\documentclass{article}

\usepackage[version=3]{mhchem} % Formula subscripts using \ce{}

\usepackage{mathrsfs}
\usepackage[margin=3cm]{geometry}
\usepackage{mathtools}
\usepackage{amsmath,amssymb,xspace,bm}
\usepackage{gensymb}
\usepackage{textcomp}
\usepackage{algorithm}
\usepackage{algpseudocode} % For algorithm writing
\usepackage{hyperref}
\usepackage[backend=biber, sorting=none]{biblatex}
\usepackage{authblk}
\usepackage{booktabs} % for clean lines
\usepackage{siunitx}  % for aligned numbers
\usepackage{soul}
\usepackage{bbm}
\DeclareSIPrePower\minus{\ensuremath{−}}

\usepackage{xcolor}

\newcommand{\bx}{\mathbf{x}}
\newcommand{\comment}[1]{}

\newcommand{\supplementarysection}{%
\setcounter{section}{0}% Reset section counter
\renewcommand{\thesection}{S\arabic{section}}% Prefix section numbers with S
 \setcounter{figure}{0}% Reset figure counter
 \let\oldthefigure\thefigure% Capture figure numbering scheme
 \renewcommand{\thefigure}{S\oldthefigure}% Prefix figure number with S
 \section*{Supporting Information}% Set supplementary section
 \let\oldsection\section% Copy \chapter into \oldchapter
 \renewcommand{\section}{% Update \chapter
 \let\thefigure\oldthefigure% Copy \thefigure into \oldthefigure
 \let\section\oldsection% Restore original \chapter
 \oldsection% Call original \ref\chapter
 }
}

\author[1,2]{Klara Suchan}
\author[1]{Shaswat Mohanty}
\author[1]{Hanfeng Zhai}
\author[1]{{Wei Cai}\thanks{caiwei@stanford.edu}}

\affil[1]{\footnotesize Department of Mechanical Engineering, Stanford University, Stanford, CA 94305}
\affil[2]{\footnotesize Department of Physics, Synchrotron Physics, Lund University, Lund}

\title{Learning Interatomic Force Coefficients from X-ray Thermal Diffuse Scattering Data}

\begin{document}
\maketitle
\begin{refsection}

\begin{abstract}
We present a fully automated framework for extracting interatomic force constants (IFCs) directly from X-ray thermal diffuse scattering (TDS) data. By formulating scattering intensity as a differentiable function of a symmetry-reduced IFC parameterization, we enable gradient-based optimization via direct, Cholesky-based sampling of correlated atomic displacements at thermal equilibrium. This approach bypasses the computational bottleneck of repeated Hessian matrix diagonalizations, significantly accelerating the inversion process. Benchmark tests demonstrate that the framework accurately recovers ground-truth IFCs and phonon dispersion relations, providing a robust, high-throughput pathway for studying lattice dynamics across diverse crystalline materials. This method bridges the gap between experimental observations and computational modeling, enabling the direct integration of TDS data into the refinement of high-fidelity interatomic potentials.
%
%We establish a fully automated framework for extracting interatomic force constants (IFCs) directly from X-ray thermal diffuse scattering (TDS) data, enabling quantitative analysis of lattice dynamics in crystalline materials, and accurate validation and refinement of interatomic potentials. We express scattering intensity as a differentiable function of a symmetry-reduced IFC parameterization, computed via direct (Cholesky-based) sampling of the correlated thermal displacements, thereby avoiding repeated diagonalizations of the Hessian and enabling automatic-gradient optimization. Our approach accurately recovers both the ground-truth IFCs and phonon dispersion relations. This framework provides a robust and flexible pathway for exploiting TDS data for the study of lattice dynamics and phonon-mediated properties across a wide range of crystalline materials.  It enables the direct incorporation of experimental data into the development of interatomic interaction models, bridging the gap between computational modeling and experimental validation.
\end{abstract}

\section{Introduction}

\begin{figure}[!htbp]
    \centering
    \includegraphics[width=0.65\linewidth]{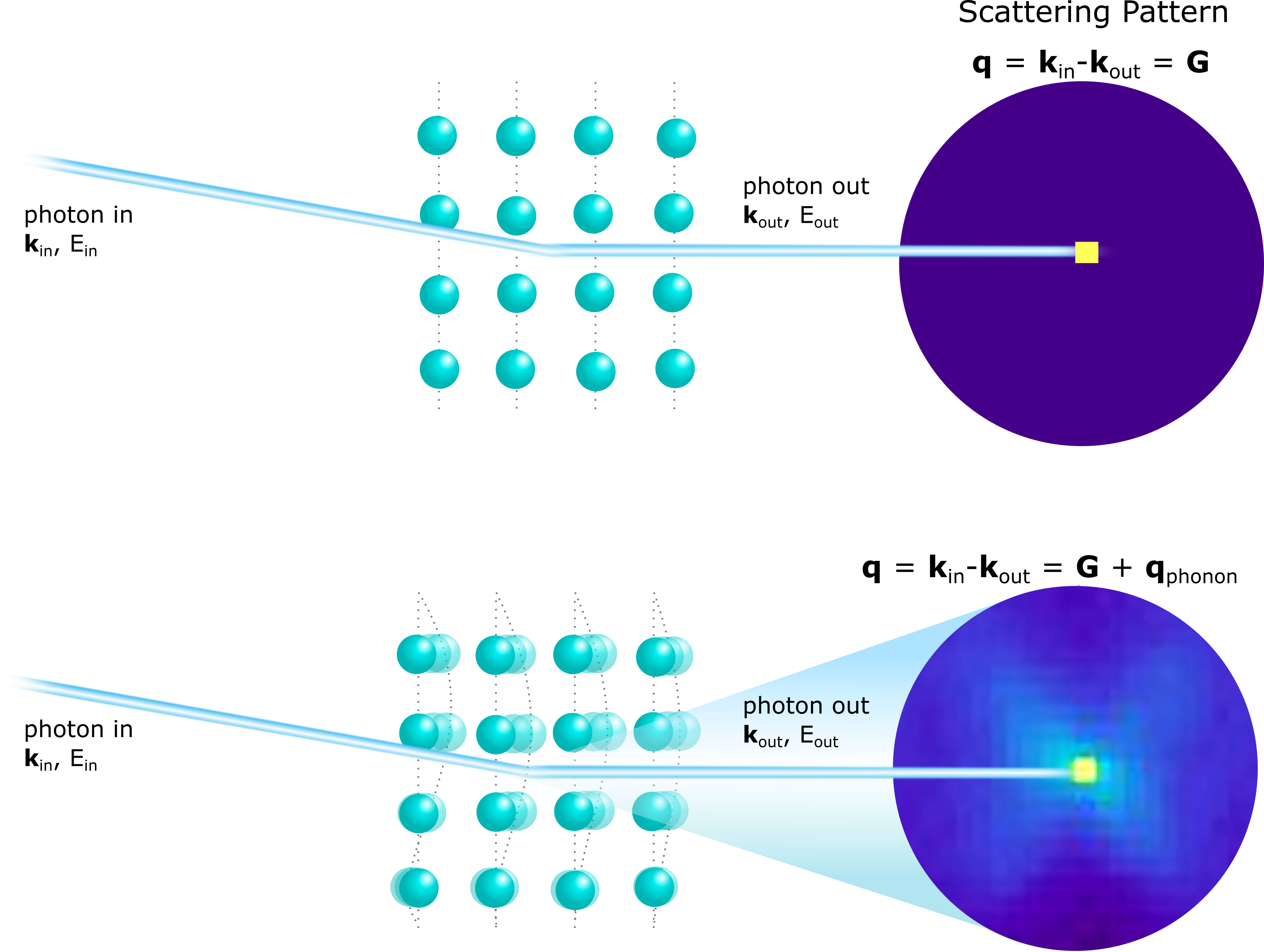}
    \caption{Schematic representation of thermal diffuse scattering. In the absence of atomic vibrations, elastic scattering leads to sharp Bragg peaks in the scattering pattern. Atomic vibrations introduce diffuse signal around the Bragg peaks due to inelastic scattering. The specific pattern of the diffuse scattering signal reflects the correlation of atomic vibrations governed by interatomic force, making TDS a direct probe of the interatomic force constants (IFCs) in materials.}
    \label{fig:sketch_TDS}
\end{figure}

%The interatomic coupling determines most material properties and phonons are a link between IFC and measurable scattering signal.
%Importance of IFCs
%
%How atoms interact with each other in a lattice underlies essentially all properties of solid-state materials, from thermal conductivity and elastic response to novel phenomena such as charge density waves and superconductivity
The interactions between atoms are fundamental to all properties of solid-state materials, ranging from elasticity and thermal conductivity to emergent phenomena like charge density waves and superconductivity~\cite{born_dynamical_1955, frohlich_theory_1997, bardeen_theory_1957, ziman_electrons_2001, giustino_electron-phonon_2017, xie_thermal_2025}.
For crystalline materials, these interactions can be formally quantified by a set of interatomic force constants (IFCs), also known as Born-von K{\'a}rm{\'a}n force constants~\cite{born_dynamical_1955, cochran_lattice_1963}. 
While IFCs are traditionally derived from computational models spanning a range of sophistication—from first-principles density functional theory~\cite{parlinski_first-principles_1997} to machine-learned empirical potentials~\cite{lee_machine_2024}—extracting them directly from experimental measurements is essential. Such direct inversion provides the definitive reference data required to validate and refine theoretical models against physical reality.
%IFCs can be obtained from computational models of interatomic interactions with various level of sophistication, from first-principles density-functional theory~\cite{parlinski_first-principles_1997} to empirical interatomic potential models~\cite{lee_machine_2024}.
%Obtaining IFCs directly from experiments provides crucial reference data for validating and refining these computational models.
%add reference to paper Does Hessian Data Improve the Performanceof Machine Learning Potentials?\cite{rodriguez_does_2025}

The significance of IFCs also stems from their direct mathematical connection to the phonon spectrum,
%There is a direct connection between the IFCs and the spectrum of phonons
 which represents the collective vibrations of the atoms in the crystal lattice.
Since phonons scatter incident radiation across a wide range of probes (e.g., X-rays, neutrons, electrons), they directly link IFCs to experimentally measurable quantities~\cite{dastuto_phonon_2002}.
%Since phonons are responsible for mediating the majority of signals observed when a material is probed by incident radiation (e.g. X-rays or neutrons)~\hl{[Ref]}, the IFCs are thus uniquely tied to experimentally measurable quantities.
%
%Therefore, through phonons, there exists a unique connection between IFCs and experimentally measureable quantities.
%
%Phonons, the collective vibrations of atoms, provide a direct link between the atomic-scale interactions and macroscopic material responses. Through phonons, IFCs uniquely connect the atomic potential energy with experimentally observable quantities. 
%
Obtaining quantitative, experimentally-derived information on IFCs is crucial for advancing our understanding of lattice dynamics, improving predictive atomistic simulations, and enabling the knowledge-driven design of new functional materials.
%
%Detailed information on IFCs would hence allow interatomic potentials to be validated against experimental insights, advancing fundamental understanding, improving predictive simulations, and enabling the knowledge-driven design of new functional materials.

%Summary of existing methods and why TDS is best
%TDS measures all phonons in entire brillouin zone (k-space) (3D) and X-TDS combines that with the flexibility of X-rays (accesible, inexpensive, can work with various samples and sample geometries.)
Despite their central role in materials physics, the accurate and systematic determination of IFCs directly from experiments remains a challenge. 
%
%IFCs can be obtained from computational models of interatomic interactions with various level of sophistication, from first-principles density-functional theory~\cite{parlinski_first-principles_1997} to empirical interatomic potential models~\cite{lee_machine_2024}, and having ways to determine IFCs from experiments provides important benchmarks to the computational models.
%
%IFCs can be obtained from first-principles density-functional theory~\cite{parlinski_first-principles_1997} or from machine learning interatomic potenials trained on ab-initio data~\cite{lee_machine_2024}.
%
Experimental access to phonon dispersions and thereby to the IFCs, is commonly sought using Raman spectroscopy, inelastic neutron scattering (INS)~\cite{nicklow_lattice_1972, neuhaus_role_2014}, and inelastic X-ray scattering (IXS)~\cite{schwoerer-bohning_phonon_1998, mohr_phonon_2007}.
However, each method has its own limitations.
Raman spectroscopy is inherently restrictive, as it probes only zone-center phonons and only those that are Raman-active, leaving most of the Brillouin zone inaccessible.
INS offers the desired full-zone phonon dispersions, but it typically requires large single crystals, involves long measurement times, and requires access to specialized, limited-availability neutron sources~\cite{dastuto_phonon_2002}. 
IXS offers data richness comparable to INS, however, it suffers from intrinsically low flux, limited reciprocal-space coverage, and point-by-point mapping, requiring extensive sample rotation and energy scans. As a result, full Brillouin-zone phonon mapping using IXS is extremely time-consuming and often impractical~\cite{dastuto_phonon_2002, baron_high-resolution_2016}.

Thermal diffuse scattering (TDS) captures the inelastic scattering occurring away from the Bragg peaks.
This scattering arises from the instantaneous atomic displacements in a thermally vibrating crystal (see Fig.~\ref{fig:sketch_TDS}).
Consequently, TDS provides a continuous map of phonon intensities across the entire reciprocal space, encompassing both acoustic and optical modes.
These unique characteristics allows the TDS to serve as a robust probe of the IFCs.
The utility of phonon scattering for validating machine-learned interatomic potentials (MLIP) has recently been demonstrated using electron TDS~\cite{cui2023machine, kim_modeling_2024,mohanty2024generalizability}.
X-ray TDS (X-TDS) extends this potential by leveraging 
%combines the potential of phonon scattering with 
the versatility and accessibility of modern X-ray facilities.
Unlike INS or IXS, X-TDS measurements can be acquired rapidly over vast regions of reciprocal space and can be readily performed under diverse \textit{in situ} conditions (e.g. varying temperature, pressure, or illumination). However, extracting information on IFC embedded in the X-TDS data is not trivial; it requires sophisticated modeling and computational inversion to decode the atomic interactions from the scattering signal.

%X-TDS is usually modeled by MD, you get atomic density for each configuration, do FFT and square to get intensity and then time-average (solves the forward problem). But it is impractical to fit interatomic interaction coefficients from TDS data using this approach (inverse problem).
%State of the art modelling
%
Computational X-TDS images are typically generated from atomic trajectories via molecular dynamics (MD) simulations~\cite{bergsma1986transient}. 
Because X-TDS is a time-averaged quantity, producing a statistically reliable image requires MD trajectories consisting of millions of time steps to sample a sufficient number of independent configurations.
This requirement for statistical rigor renders the direct MD approach computationally prohibitive, particularly when address the inverse problem: the quantitative extraction of IFCs from experimental data.
%Obtaining sufficient statistically independent configurations for a time-averaged image requires trajectories of millions of steps, making it computationally expensive. 
%
%The most significant difficulty arises when attempting to solve the inverse problem: the quantitative extraction of material parameters, such as IFCs, from experimental X-TDS data. 
%
An efficient inversion would typically require a gradient-based optimization algorithm, which requires calculating the gradients of the X-TDS intensity with respect to the IFC parameters. Using automatic differentiation (autograd) in this context would necessitate back-propagating derivatives through a multi-million-step trajectory, which is computationally infeasible.
%
%This fitting process requires calculating the gradients of the X-TDS image with respect to the IFCs. 
%
%Employing automatic differentiation (autograd) would necessitate the backward propagation of derivatives through the entire multi-million step trajectory, rendering it computationally infeasible.
%
%Furthermore, the computation of gradients with respect to interatomic force constants would have to be calculated through the complete trajectory, consisting of millions of steps, making it infeasible. 
%
Alternatively, estimating the gradient via the finite-difference method introduces significant numerical noise and scales poorly with the number of IFC parameters.
%
%Finite-difference gradient estimation in turn, introduces noise and scales poorly with the system size.
%
As a result, MD-based X-TDS simulations are primarily confined to generating images for qualitative comparison with experiments, making the method unsuitable for the task of quantitative extraction of IFCs from experimental X-TDS data.
%
%As a result, MD-based X-TDS simulations are primarily used for the computation of single TDS images and qualitative comparison to experiments. The computational expense associated makes it undesirable for solving the inverse problem and doing a quantitative parameterization of interatomic potentials. 

%******************** 2025/11/20 ********************

%Quantitative X-TDS (less commonly used, trying to solve the inverse problem) - by diagonalization of Hessian matrix - use analytic expressions from eigenvectors to TDS - limited to first order - (someone else did higher order - add reference here) (We don't diaganize Hessian)
To address the inverse problem, a quantitative X-TDS analysis has been developed by \textcite{holt_determination_1999, holt_x-ray_2001, holt_phonon_2002}.  This approach is based on an analytical derivation of the scattering intensities from the eigenvalues of the Hessian matrix. It has been demonstrated for selected model materials like Ni, Si, Nb, PuGa TiSe2, and MgO~\cite{holt_determination_1999, holt_x-ray_2001,  holt_phonon_2002, xu_determination_2005,wong_intensity_2016, mei_reflection_2015}. However, this eigenvalue-based approach faces intrinsic limitations: it requires repeated diagonalization of the dynamic matrix, resulting in a computationally costly cubic scaling with system size.
Furthermore, the model treats phonon modes as independent entities, which complicate the inclusion of higher-order scattering effects (like phonon-phonon scattering) that can contribute up to $40 \%$ of the total intensity at elevated temperatures~\cite{barabash_diffuse_2009, xu_determination_2005}.
%
%To overcome these challenges, 
While \textcite{wehinger_full_2017} developed an alternative approach to obtain the elastic tensor from the TDS successfully, 
%. While this is successful in obtaining the elastic tensor, 
it has not led to the extraction of the entire interatomic force constant matrix.
Collectively, these limitations have prevented X-TDS from becoming a routine quantitative tool for recovering IFCs. 
Despite its remarkable experimental accessibility and information content, experimental TDS images are thus largely analyzed qualitatively, by visually comparing broad $q$-space trends to other experimental methods such as INS or computational X-TDS images~\cite{weadock_nature_2023}.

%A fast and general method to extract IFCs from X-TDS would allow the measurement of phonons and validation of interatomic potentials (more general than that? extract interatomic interaction coefficients, extract fundamental physical parameters).
A fast, robust, and general method capable of extracting IFCs from X-TDS data would 
provide direct experimental access to the fundamental forces governing atomic interactions in solids.
This information is essential to the understanding of a material's intrinsic dynamics and response to external stimuli.
%response to temperature, stress, and other stimuli.
%
%The resulting quantitative information would be highly valuable to provide atomic-level validation of theoretical models and interatomic potentials, while enabling quantitative characterization of phonons across the entire Brillouin zone.
%
%Our approach overcomes previous limitations by using the crystal symmetry and directly computing uncorrelated atomic configurations from the cholesky decomposition of the hessian allowing for a fully differentiably framework.
%In this paper, ...
To enable the direct extraction of the IFCs from X-TDS data, we develop a fully differentiable end-to-end framework connecting IFCs to the X-TDS images. Our approach combines and builds on three key advances.
First, we develop an algorithm to utilize the crystal's space group symmetry to construct a symmetry-reduced Born-von K\'{a}rm\'{a}n parametrization of the Hessian, yielding a minimal and physically consistent set of independent IFCs. These ensure both physical consistency and reduction of computational complexity. Second, we evaluate the TDS without the phonon eigen-modes by directly sampling correlated thermal displacements from the Hessian via a differentiable Cholesky factorization and computing scattering intensities using FFT-based methods.  Third, this differentiable forward model enables automatic gradient computation with respect to the IFCs.
This strategy eliminates the need for repeated dynamical-matrix diagonalizations and enables efficient, gradient-based refinement of interatomic force constants from X-TDS.
Furthermore, this method paves the way for routine, quantitative X-TDS analysis for both fundamental phonon studies and data-driven materials modeling~\cite{xu_determination_2005, benitez_why_2025}.

%Structure of the remaining paper.
The remainder of this paper is structured as follows. In Section~\ref{sec:methods}, we describe our computational method, including the complexity reduction techniques using crystal symmetry, the direct sampling approach for generating atomic displacements, and the resulting gradient descent algorithm for optimizing the IFCs. 
Section~\ref{sec:results} presents the results of our method applied to benchmark cases, demonstrating convergence and accuracy in IFC extraction.
%the parameterization of the Hessian to the thermal diffuse scattering signal. 
%
The implications of the results and limitations of the approach are discussed in Section~\ref{sec:Discussion}.
Finally, in Section~\ref{sec:conclusions}, we summarize our key findings and outline potential avenues for further development and applications for this computational framework.

\section{Methods} \label{sec:methods}
%Overview of the framework
%How we solve the forward problem - calculate TDS from interatomic force coefficients - We overcome the limitations by bypassing the eigenvalue problem. We use symmetry, cholesky decompostions to calculate atomic displacements that fullfill physical correlations and then to a FFT to obtain the TDS without the need to diagonalize the hessian matrix.
%Why now the inverse problem is possible to be solved - every step in the forward problem is 'differentiable' and fast enough, as a result, the 'error' can be 'back-propagated' to the parameters
To enable efficient and quantitative extraction of the IFCs directly from measured TDS data, we reformulate the forward scattering model (from the IFCs to TDS images) to be both fully differentiable and computationally efficient.
This differentiability is the key innovation: it allows errors between simulated and target intensities to be back-propagated to enable gradient-based optimization.
%of the underlying force constants. 

%We represent the interatomic interactions directly through the Born–von Kármán force constants, which constitute the symmetry-dictated minimal set of free parameters. These are mapped into the real-space Hessian matrix, exploiting the material's space group symmetries. From this Hessian matrix, atomic displacements are sampled in a differentiable manner, which inherently fulfills the physical correlations dictated by the IFCs. 
%The diffuse scattering intensity is then computed directly from the atomic configurations obtained from the sampled displacements, using a fast Fourier transform (FFT) approach, developed by \textcite{Mohanty2022}.
%This results in an end-to-end differentiable framework, which retains the full physical content of the harmonic model. This hence enables gradient-based optimization to extract the IFCs by minimizing discrepancies between simulated and target intensities.
%The framework is illustrated in Fig.~\ref{fig:Sketch_method}.

\begin{figure}[!htbp]
    \centering
    \includegraphics[width=0.95\linewidth]{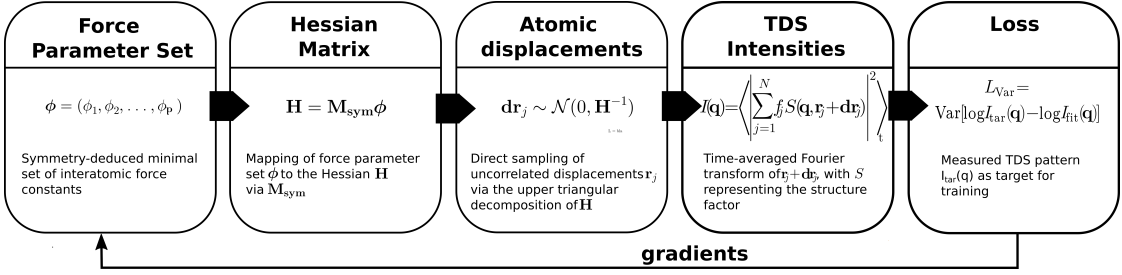}
    \caption{
    Schematic of the differentiable optimization loop. The forward pass (bold arrows) generates predicted TDS intensities through stochastic displacement sampling and FFT-based scattering calculations. During the backward pass (thin arrow), gradients of the loss function are propagated to update the parameters $\phi$ iteratively.}
    %Schematic illustrating the optimization loop. The differentiable forward bath (bold arrows) generates predicted TDS intensities via displacement sampling and FFT-based scattering calculations. Gradients flow backward (thin arrow) to update $\phi$ via the computed loss.
    %Schematic of the gradient descent framework for extracting force constants from TDS. A symmetry-reduced set of force constants ($\phi$) defines the Hessian matrix H, from which correlated atomic displacements are sampled. TDS intensities are computed by averaging the scattering signal over independent displacement configurations. 
    %The loss is defined from the difference between the computed and target scattering signals.
    %Differentiability of the entire forward calculation enables automatic gradient calculation of the loss with respect to $\phi$, allowing direct optimization via gradient descent.
    \label{fig:Sketch_method}
\end{figure}

The optimization workflow is illustrated in Fig.~\ref{fig:Sketch_method}.
We start by initializing the values of a symmetry-reduced minimal set of IFCs ($\boldsymbol{\phi}$) that describes the atomic interactions within a chosen cut-off radius.
This minimal set is then used to explicitly and linearly construct the entire Hessian matrix ($\mathbf{H}$) for a crystal supercell through an inherently differentiable mapping.
The next critical step involves efficiently and differentially generating a set of random atomic displacements that satisfy the Boltzmann distribution dictated by the Hessian matrix.
This is accomplished through a Cholesky decomposition of the Hessian matrix, which ensures both efficiency and differentiability.
Subsequently, the X-ray scattering intensity image is computed for each generated atomic configuration using the Fast Fourier Transform (FFT) approach~\cite{Mohanty2022}.
The average of these intensities yields the computational prediction $I(\mathbf{q})$ for the X-TDS image corresponding to the current IFC values.
This predicted image is then compared with the experimentally measured X-TDS image, and the loss is computed as the root mean square error.
The derivatives of the loss function with respect to the IFCs are calculated through automatic differentiation (\texttt{autograd}).
Finally, a gradient-descent algorithm uses these derivatives to update the IFC values.  
This entire cycle constitutes one epoch of the training process, which is typically repeated for thousands of epochs until the IFCs converge to the optimal target values.

\subsection{Minimal IFC Set} \label{subsec:symmetry}
%Hessian - (instead of the full interatomic potential) - (discussion: generalization) (e.g. normalizing flow - generate one-shot samples of Boltzman distribution fast and differentiable)
%Problem - Hessian matrix has too many elements - but they are related by symmetry - irreducible elements are much fewer
%By exploiting crystal symmetry and physical constraints, the enormous number of interatomic force constants in a crystal can be reduced to a minimal, symmetry-constrained set of independent parameters that fully describe the material’s lattice dynamics.

%******************** 2025/11/21 ********************

As a concrete example, we consider computational X-TDS from an atomistic model of face-centered cubic (FCC) Ni.
We choose a supercell of size $5[1\,0\,0]\times 5[0\,1\,0]\times5[0\,0\,1]$, which contains $N = 500$ atoms and is subjected to periodic boundary conditions in all three directions.
%
%\hl{(Why do we choose this supercell size?  What happens if the number of atoms is too small?)}

This large supercell size is chosen to minimize finite-size effects. A smaller simulation cell results in a much coarser resolution in the $\mathbf{q}$-space over which the X-TDS images will be computed.
Let $U(\{x_i\})$ be the interatomic interaction potential function we wish to learn from measured TDS images, where $\bx\equiv\{x_i\}$, $i=1\,\cdots,3N$ is the array of all atomic coordinates.
We can Taylor expand $U(\{x_i\})$ around the perfect crystal positions $x^0_i$, for which the forces on all atoms are zero.
\begin{equation}
 U(\{x_i\}) = U(\{x^0_i\}) + 
  \frac{1}{2} \sum_{i,j} H_{ij} 
  \, \delta x_i \, \delta x_j
  + \mathcal{O}(\{\delta x_i\})^3 \, ,
\end{equation}
where $\delta x_i = x_i - x_i^0$ is the atomic displacements from the perfect crystal positions and $H_{ij}$ are the second-order partial derivatives of the potential,
\begin{equation}
H_{ij} \equiv \left.\frac{\partial^2 U}
{\partial x_i \partial x_j} \right|_{\{x_i = x_i^0\}} \, .
\end{equation}
At temperatures much lower than the melting temperature, the atomic displacements $\{\delta x_i\}$ are sufficiently small that the higher-order terms, $\mathcal{O}(\{\delta x_i\})^3$, in the Taylor expansion are negligible.
By truncating the expansion after the second-order term, we employ the harmonic approximation (HA), a standard approach in TDS analysis.
%
%don't think this is strictly true
The coefficients $H_{ij}$ are also called the Born-von K\'{a}rm\'{a}n force constants.  These coefficients collectively form the $3N\times 3N$ Hessian matrix $\mathbf{H} = \{ H_{ij} \}$.
This Hessian matrix completely describes the interatomic forces within the HA. The central goal of our TDS image analysis is to learn the force constants, hence the Hessian matrix, from the experimental data.

A first challenge one encounters in learning the Hessian matrix is the large number of its components.
In our current example of $N = 500$ atoms, the Hessian matrix contains $2,250,000$ components, which is too many to deal with.
Fortunately, the number of independent matrix components is drastically smaller due to the symmetry inherent in the crystal structure.

First, FCC Ni has a simple crystal structure, with only one atom per unit cell. Consequently, all atoms are symmetry-equivalent due to the crystal's translational symmetry.
This means the relationship between any two atoms, $k$ and $l$, is identical to the relationship between any other two atoms, $m$ and $n$, which are separated by the same lattice vector.
Specifically, the translational symmetry implies that the $3\times 3$ sub-block of the Hessian matrix relating atoms $k$ and $l$ only depends on the difference in the equilibrium positions between these two atoms.
Therefore, if we determine the force constants between the first atom (atom 0) and all other atoms in the supercell, we can construct all the remaining components of the Hessian matrix.
This allows us to focus our learning efforts exclusively on the $9N$ components of the first three rows of the Hessian matrix, $\mathbf{H}_{ij}$ (where $i=0, 1, 2$, $j=1, \dots, 3N$), which contains $4,500$ components.

%In other words, every atom is equivalent to each other due to the translational symmetry of the lattice.
%
%This means that if we obtain the first three rows $(i=0,1,2)$ of the Hessian matrix (corresponding to the first atom), we can construct all the remaining rows of the matrix.
%
%Hence in the following we will focus on the matrix components $H_{ij}$ where $i = 0, 1, 2$ corresponding to the first atom (which we choose to be located at the origin). 

%Atoms in a solid are coupled via the potential energy $U$. Rather than parameterizing the full interatomic potential, here we work within the harmonic approximation, which is valid for small displacements around equilibrium positions. The main components of the framework nonetheless can be generalizable beyond that approximation, as will be discussed in Section~\ref{sec:Discussion}. In the harmonic approximation, atomic interactions can be described by the Hessian matrix, the second derivative of the potential energy $U$ with respect to atomic displacements. However, the Hessian matrix for an exemplary FCC lattice, and a medium-sized $5\times 5 \times 5$ supercell ($N$ = 500 atoms) contains  $2.25\times 10^6$ elements, which is computationally prohibitive to parameterize directly.
%
%Even imposing a cut-off radius to include only the 5 nearest neighbor shells, reduces the parameter space  to $\mathcal{O}(N)$, yet still amounts to $54756$ parameters.

%FCC is a simple crystal structure.  The basis consists of a single atom, meaning every atom is equivalent.
%

\begin{figure}
    \centering
    \includegraphics[width=0.25\linewidth]{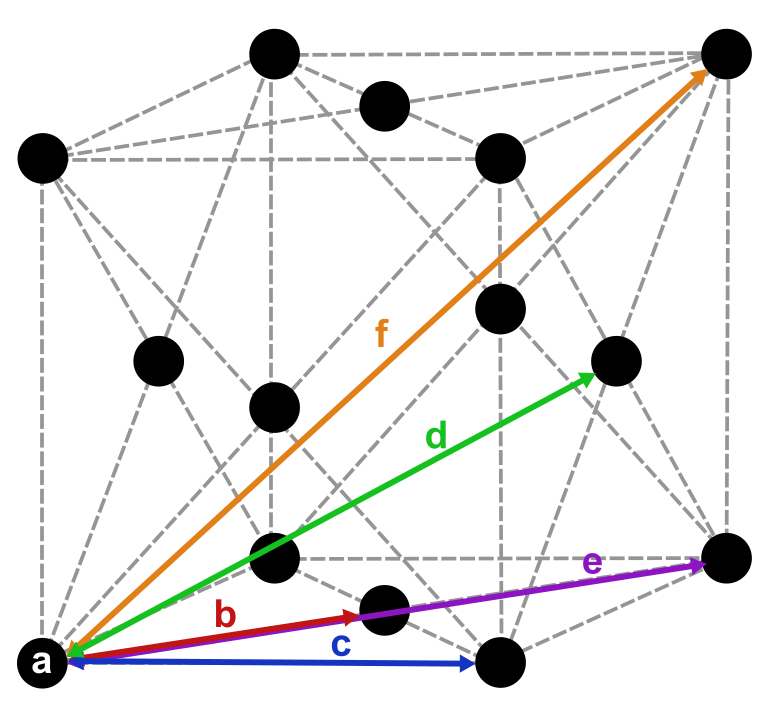}
    \caption{
    %Schematic illustration of the types of interatomic interactions corresponding to IFCs within 5-th neighbors. The ``0-th shell'' coefficient $a_0$ corresponds to the self-interaction in the Hessian matrix.  The 1st–5th shell coefficients $b_{i}$–$f_{i}$ correspond to interaction between an atom and its successive neighbors.
    %
    %Schematic of interatomic force constants (IFCs) mapped by coordination shell. The term $a_0$ represents the on-site self-interaction (diagonal elements of the Hessian matrix), while terms $b_k$ through $f_k$ denote interactions between the reference atom and its first through fifth nearest-neighbor shells, respectively.
    %
    Illustration of the IFC hierarchy for the first five neighbor shells. The 0-th shell ($a_0$) corresponds to the self-interaction term of the Hessian matrix. Coefficients $b_k, c_k, d_k, e_k,$ and $f_k$ correspond to the force constants for the 1st, 2nd, 3rd, 4th, and 5th coordination shells, respectively.
    }
    \label{fig:IFC_shells}
\end{figure}

Without loss of generality, we choose the perfect crystal position of atom 0 to be the origin of the coordinate system.
To systematically organize the force constants, we sort all atoms into shells based on their distance from atom 0.
For example, atom 0 itself is referred to as shell 0.
The 12 nearest neighbor atoms of atom 0 form shell 1; the 6 second nearest neighbor atoms form shell 2, etc.
We expect the strength of the interatomic interaction, hence the magnitude of the Hessian matrix elements $H_{ij}$, to rapidly decay as the distance between the corresponding atom $k$ and atom 0 increases.
This justifies the use of a cut-off distance beyond which the force constants are approximated as zero.
In this example, we assume all Hessian matrix elements to be zero beyond the fifth nearest neighbor shell.  As a result, the number of Hessian matrix elements that we still need to consider is now limited to 711.
%\hl{711 (confirm this)}.
%
We labeled the force constants as $a_k$, $b_k$, $c_k$, $d_k$, $e_k$, $f_k$, corresponding to shell 0, 1, 2, 3, 4, 5, respectively.

The matrix elements corresponding to the same shell are related to each other through the rotational symmetries of the crystal's point group.
For the FCC crystal structure, the corresponding point group is ${\rm O}_{\rm h}$.
For example, the $3\times3$ sub-block of the Hessian matrix, corresponding to shell 0, represents the restoring force when atom 0 is displaced.
The ${\rm O}_{\rm h}$ group requires that this matrix to be isotropic, i.e. a scalar times the identity matrix.
Hence, the 9 matrix elements for shell 0 contain only one independent constant, $a_0$.
It can be shown that there are only three independent constants, $\{b_0, b_1, b_2\}$, for shell 1; two independent constants, $\{c_0, c_1\}$, for shell 2; four independent constants, $\{d_0, d_1, d_2, d_3\}$ for shell 3; three independent constants, $\{e_0, e_1, e_2\}$ for shell 4; and four independent constants, $\{f_0, f_1, f_2, f_3\}$.
Thus the point group ${\rm O}_{\rm h}$ reduced the number of independent force constants to 17.\cite{frei_direct_1985}

There is one more symmetry satisfied by the Hessian matrix beyond the ones considered above.
Because the potential energy $U$ only depends on the relative positions of the atoms with respect to each other, no force is produced on any atoms if all atoms are subjected to a rigid-body translation from their perfect crystal positions.
As a result, every row of the $H_{ij}$ matrix sum up to zero.
In the FCC example considered here, this leads to one more constraint on the force constants.
\begin{equation}\label{eq:a0_force_constants}
%a_0 + 8 b_0 + 4 b_1 + 0 b_2 + 2 c_0 + 4 c_1 + 8 d_0 + 16 d_1 + 0 d_2 + 0 d_3 + 8 e_0 + 4 e_1 + 0 e_2 + 8 f_0 + 8 f_1 + 8 f_2 + 0 f_3 = 0
a_0 + 8 b_0 + 4 b_1 + 2 c_0 + 4 c_1 + 8 d_0 + 16 d_1 + 8 e_0 + 4 e_1 + 8 f_0 + 8 f_1 + 8 f_2 = 0
\, .
\end{equation}
%
%From SI: $a_0 =-(8 b_1+4b_2+0b_3+2c_1+4 c_2 +8 d_1+16d_2+ 0d_3+ 0d_4  +8e_1  +4e_2   +0e_3 +8f_1 +8 f_2 +8 f_3+0 f_4)$
In other words, we can always express $a_0$ as a linear combination of the other coefficients.
Thus, the number of irreducible IFCs for FCC crystal up to the fifth nearest neighbors is finally reduced to 16\cite{enkhtor_calculation_2016}; they are
\begin{equation}
    \boldsymbol{\phi} = 
    \{ b_0, b_1, b_2, c_0, c_1, d_0, d_1, d_2, d_3, e_0, e_1, e_2, f_0, f_1, f_2, f_3\} \, .
\end{equation}
These are the set of parameters we aim to learn from measured TDS images.
Once obtained, they fully determine the phonon dispersion relations as well as many other properties of the crystal, such as all the elastic constants.

We have implemented the above symmetry considerations into a Python program using the \texttt{SymPy} symbolic manipulation package.
%
%
%We developed a Python program to automatically identify the irreducible set of IFCs from the symmetry of FCC crystal using the \texttt{SciPy} package.
%
The program takes the number of nearest neighbor shells to be considered as input, and correctly deduces the set of irreducible IFCs (16 IFCs for up to 5 nearest neighbors).
It also constructs the mapping tensor $\mathbf{M}_{\rm sym}$ that constructs the full Hessian matrix $\mathbf{H}$ from the set of IFCs,
\begin{equation}
    \mathbf{H} = \mathbf{M}_{\rm sym}
       \cdot \boldsymbol{\phi} \, .
\end{equation}
By construction, the mapping from $\boldsymbol{\phi}$ to the Hessian matrix $\mathbf{H}$ is differentiable.
%
%where $\boldsymbol{\phi} = \{b_0,b_1,b_2,c_0,c_1,d_0,d_1,d_2,d_3,e_0,e_1,e_2,f_0,f_1,f_2,f_3\}$

%(\hl{I see 16 components, not 17}.)

\subsection{Sampling of Atomic Configurations} \label{subsec:direct_sampling}
%X-TDS is the time averaged scattering from a dynamic lattice and hence a TDS calculation is done on a large number of uncorrelated atomic configurations making trajectory based calculations of atomic dynamics unfeasible, due to their inherent correlation.
%Goal: sample atomic configurations, in a way that is differentiable and efficient

The full Hessian matrix $\mathbf{H}$ constructed from the set of irreducible IFCs uniquely defines a computational X-TDS image, $I(\mathbf{q})$.
Given that the typical detector exposure times far exceed the atomic vibration correlation times, the measured X-TDS image corresponds to the time-averaged diffraction signals from many atomic configurations.
Let $I(\mathbf{q}; \bx)$ be the X-ray diffraction signal from a specific atomic configuration $\bx$; 
$I(\mathbf{q}; \bx)$ can be computed from $\bx$ using a fast Fourier Transform algorithm from the corresponding atomic density field~\cite{Mohanty2022}.
In the language of statistical mechanics, the X-TDS image $I(\mathbf{q})$ is the ensemble average of $I(\mathbf{q}; \bx)$, where $\bx$ satisfies the canonical distribution at the given temperature $T$ with probability density,
\begin{equation}
    f(\bx) = \frac{1}{Z} \exp \left[ 
    -\frac{U(\bx)}{k_{\rm B}T} \right] \, .
\end{equation}
where $Z = \int_{\mathbb{R}^{3N}} \exp \left( 
    -\frac{U(\bx)}{k_{\rm B}T} \right)\,d\bx$ is the partition function.
Within the HA, 
%$f(\bx)$ is equivalent to a Gaussian distribution,
%
\begin{equation}
  f(\bx) \propto \exp \left[ 
    -\frac{\frac{1}{2} (\bx-\bx^0)^{\rm T}
     \cdot \mathbf{H} \cdot
     (\bx-\bx^0) }{k_{\rm B}T} \right] \, .
     \label{eq:gaussian_HA}
\end{equation}
This means that $f(\bx)$ is equivalent to a multivariate Gaussian distribution, $\bx \sim \mathcal{N}(\boldsymbol{\mu},\boldsymbol{\Sigma})$, with mean $\boldsymbol{\mu} = \bx^0$ and variance $\boldsymbol{\Sigma} = k_{\rm B} T \cdot \mathbf{H}^{-1}$.
In our approach, we compute an estimate of $I(\mathbf{q})$ by averaging $I(\mathbf{q};\bx)$ over a set of  $N_{\rm s}$ atomistic configurations sampled over the Gaussian distribution, Eq.~(\ref{eq:gaussian_HA}),
\begin{equation}
I_{\rm comp}(\mathbf{q}) =
\frac{1}{N_{\rm s}} \sum_{n=1}^{N_{\rm s}}
I(\mathbf{q}; \bx^{(n)}) 
\end{equation}
To do so requires the generation of samples from the distribution, Eq.~(\ref{eq:gaussian_HA}).

To enable gradient-based optimization of the IFCs, this sampling step must be accomplished in a way that allows efficient computation of the derivative of the result with respect to the Hessian matrix.
We accomplish this by performing the Cholesky decomposition of the Hessian matrix,
\begin{equation}
\mathbf{H} = \mathbf{L} \cdot 
 \mathbf{L}^{\rm T} \, .
\end{equation}
where $\mathbf{L}$ is a lower triangular matrix.
Recall that the atomic displacement $\delta \bx = \bx - \bx^0$ satisfies the Gaussian distribution, $\delta \bx \sim \mathcal{N}(\mathbf{0}, k_{\rm B}T\,\mathbf{H}^{-1})$.
Let $\mathbf{u} \sim \mathcal{N}(\mathbf{0}, \mathbf{I})$ be a random variable satisfying the standard Gaussian distribution with zero mean and identity matrix as the variance.
The atomic displacement can be obtained by solving the matrix equation,
\begin{equation}
    \mathbf{L}^{\rm T} \cdot \delta\mathbf{x} = \sqrt{k_{\rm B}T} \, \mathbf{u} \, .
    \label{eq:L_dx_u}
\end{equation}
%
%\hl{(Please verify that the code has the $\sqrt{k_{\rm B}T}$ term.)}
%
For each random sample of $\mathbf{u}^{(n)}$, an independent atomic configuration can be obtained by solving Eq.~(\ref{eq:L_dx_u}), and then adding the displacement to the perfect crystal positions, $\bx^{(n)} = \bx^0 + \delta \bx^{(n)}$.
By implementing the above steps in \texttt{PyTorch}, the derivatives of the atomic configuration $\bx^{(n)}$ with respect to the Cholesky factor $\mathbf{L}$ and then to the Hessian matrix $\mathbf{H}$ can be computed by backward propagation.

\subsection{IFC Optimization} \label{subsec:optimization}
%Optimazation of IFCs is done by gradient descent fit with loss being the difference of the log of intensities and an additional penalty loss if the hessian become not positive definate, enabling robust optimization of IFCs.

The methods described above are combined to give an efficient algorithm that takes an irreducible set of IFCs $\boldsymbol{\phi}$ as input, and outputs the X-TDS images $I_{\rm fit}(\mathbf{q})$ over a 3-dimensional grid of reciprocal space vectors $\mathbf{q}$.
Importantly, the algorithm also computes the derivatives of the output $I_{\rm comp}(\mathbf{q})$ with respect to the input $\boldsymbol{\phi}$ efficiently through backward propagation.
Given a target X-TDS image, $I_{\rm tar}(\mathbf{q})$, e.g., from experiments, the goal is to search for the IFC values, $\boldsymbol{\phi}$, that minimize the variance of the difference between the logarithm of the computed and target X-TDS images.
%
%a gradient-based optimization algorithm can be constructed to search for the IFC values that minimize the root mean square (RMS) error between the computed and target X-TDS images.
%
This variance serves as the primary loss function, and is defined as
%
%\begin{equation}
%    L_{\rm RMS} = \sqrt{\frac{1}{N_{\rm q} } \sum^{N_{\rm q}}_{s=1} 
%      \left[ \log I_{\rm tar}(\mathbf{q}_{s}) - \log I_{\rm fit}(\mathbf{q}_s) - C \right]^2} \, ,
%      \label{eq:L_RMS_old}
%\end{equation}
%
\begin{equation}
    L_{\rm Var} = \operatorname{Var} \left[ \log I_{\rm tar}(\mathbf{q}) - \log I_{\rm fit}(\mathbf{q}) \right]_{\mathbf{q} \in \mathbf{Q}} \, ,
      \label{eq:L_RMS}
\end{equation}
%
%\hl{(Insert an adjustable constant $C$ in the loss function, because we are only interested in matching the shape of the intensity, not the absolute value? Wait for results of this approach.  Change Fig.s~2 if necessary.)}
%
%where $N_{\rm q}$ is the number of $\mathbf{q}$-space points at which the images are compared.
where $\mathbf{Q}$ is the set of $\mathbf{q}$-space points at which the images are compared.
The use of the $\log$ function in Eq.~(\ref{eq:L_RMS}) is essential.  This is because the signal containing the information about atomic motion primarily resides in the $\mathbf{q}$-space between the intense Bragg peaks. The X-ray intensity in such spaces is orders of magnitude smaller than that at the Bragg peaks. The minimization of the variance of intensity on a logarithmic scale effectively applies a homoscedastic transformation on the signal's variance over the range of the data, contributing more meaningfully to the total error, which would otherwise be dominated by the Bragg peak.  
%A logarithmic scale ensures that these regions contribute meaningfully to the total error, which would otherwise be dominated by the Bragg peak regions.

For a physically stable crystal structure, the IFC array $\boldsymbol{\phi}$ must result in a Hessian matrix $\mathbf{H}$ that is positive semi-definite.  If this condition is not met, the Cholesky decomposition of $\mathbf{H}$ cannot be completed, and the rest of the workflow cannot proceed.
To enforce this constraint, we introduce a penalty term, 
\begin{equation}
 L_{\rm penalty} = \sum _i \max(0,-\lambda _i) \, ,
\end{equation}
where $\lambda_i$ are the $3N$ eigenvalues of the Hessian matrix.
$L_{\rm penalty} = 0$ if all eigenvalues are non-negative, i.e. when $\mathbf{H}$ is positive semi-definite.
The final loss function for the optimization algorithm is a combination of the variance error and the penalty term,
\begin{equation}
L(\boldsymbol{\phi}) = L_{\rm Var} \cdot \mathbbm{1}(L_{\rm penalty} = 0)
 + \alpha \cdot L_{\rm penalty} \, ,
\end{equation}
where $\alpha$ is a hyper-parameter controlling the strength of the penalty, and $\mathbbm{1}(\cdot)$ is the indicator function, which equals 1 if the condition inside is true and 0 otherwise.
This loss function is implemented in \texttt{PyTorch} and the Adam algorithm is used to minimize the loss and optimize the IFCs.

\section{Results} \label{sec:results}
% Using randomly perturbed FCC Nickel as a benchmark, the method can accurately IFCs up to the 5th neighbor and the phonon dispersion relation, showing that vibrational dynamics encoded in scattering images are sufficient to obtain accurate interatomic interactions.

%\subsection{Distinguishing Interatomic Potentials via Thermal Diffuse Scattering}

%\subsetion{Better title}

\begin{figure}[ht]
    \centering
    \includegraphics[width=0.95\linewidth]{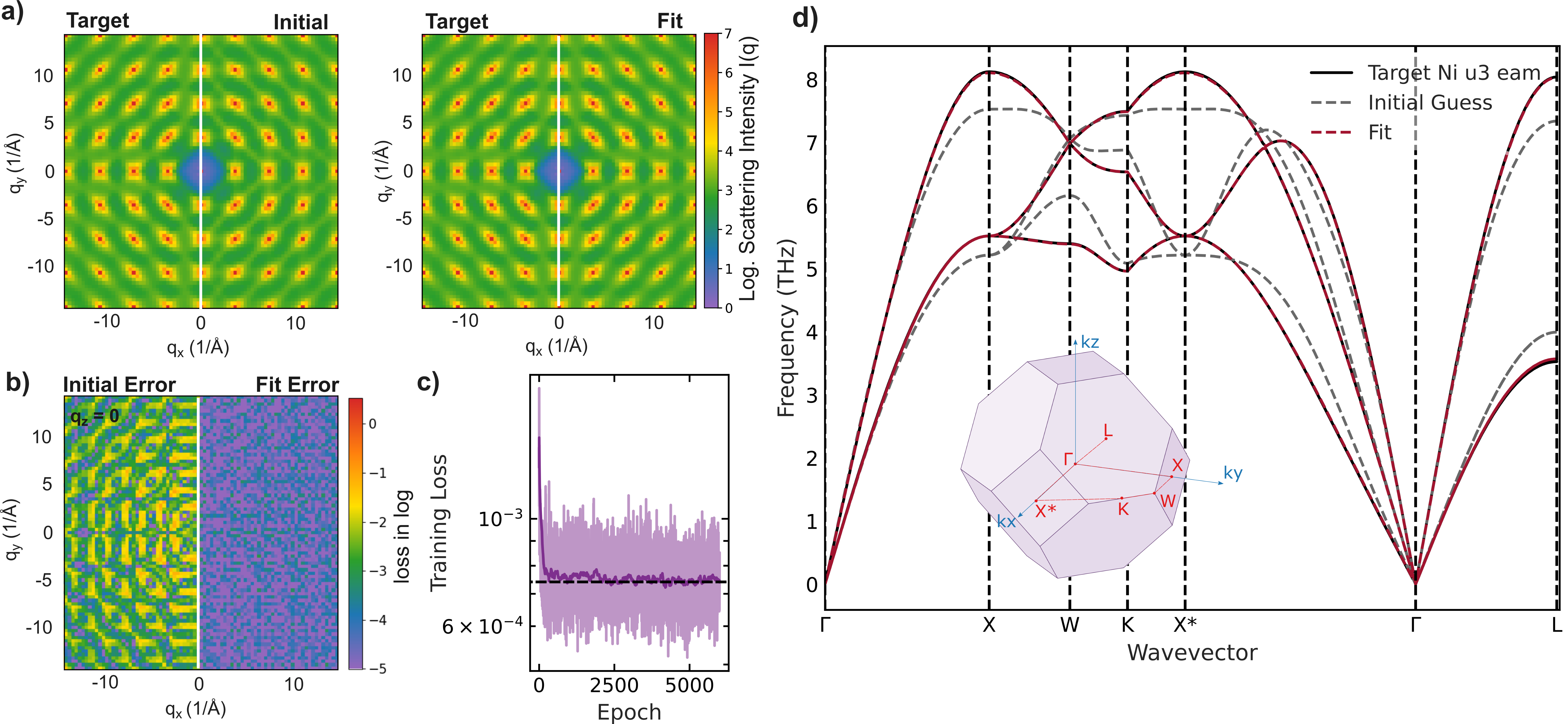}
    \caption{a) Comparison of the computed TDS signals at 300~K on the plane of $q_z = 0$ (a.k.a. the HK(L=0) slice in the reciprocal space).  The sections marked as ``Target'' correspond to the U3 EAM potential.  The section marked as ``Initial'' corresponds to the MEAM potential.  The section marked as ``Fit'' corresponds to the IFCs after the training.
    %
    %
    %, shown exemplarily for the HK0 slice. 
    b) The error of the computed TDS signals based on the initial and fitted IFCs when compared with the target.
    %The error to the target is shown to be reduced after fitting. 
    %
    c) The training loss (light purple) is shown together with the moving average (dark purple) and the final value (dashed). 
    d) The phonon dispersion relation computed from the Target, Initial, and Fitted IFCs.
    %shows significant deviations between the U3 EAM potential and the MEAM potential.
    %
    %The fitted IFCs is shown to reproduce the phonon dispersion relation of the Target accurately. 
    %
    The curves corresponding to the Target and Fitted IFCs are nearly indistinguishable here. 
    The inset shows the first Brillouin zone with the paths connecting high symmetry points shown in red.}
    %
    %(3D data to fit, ref X-TDS from 1000 configs, after 6000 epochs) 

    \label{fig:MEAM}
\end{figure}

% (first we discuss things common to Section 3.1 and 3.2)
Here we demonstrate that our method can successfully extract IFC values from TDS data.
As a benchmark, we first compute the target TDS signals for Ni modeled by an embedded-atom method (EAM) potential~\cite{foiles1986embedded} at 300~K, and use this TDS data as ground truth to extract the IFC values.
This approach allows us to compare the extracted IFC values with the ground truth, and quantify the error of the numerical approach.
%Here we present a benchmark to verify whether our method works -- for this purpose, we take EAM model for Ni as the reference, compute its TDS at 300~K, and use it to train the IFCs -- this way we can compare the trained results with the ground truth -- measure the error.
The target IFC values are obtained from the Hessian matrix of the embedded-atom method (EAM) potential evaluated at the perfect crystal configuration, with the lattice constant set to its equilibrium value at 300 K ($a = 3.52$ \AA ~\cite{hwang1972thermal}), corresponding to the quasi-harmonic approximation.
The target TDS signal used for learning the IFC values is generated from an atomistic model of a face-centered cubic (FCC) crystal of Ni of size $[5\,0\,0] \times [0\,5\,0] \times [0\,0\,5]$, containing 500 atoms, and subjected to periodic boundary conditions (PBC) in all three directions.
%
% The simulation cell is then equilibrated at $300$~K using molecular dynamics (MD) simulations with the cell dimensions adjusted iteratively to achieve zero pressure isotropically. We then minimize the energy by constraining the simulation cell volume to obtain the equilibrium atomic configuration at $300$~K.
%
The target TDS signal $I_{\rm tar}(\mathbf{q})$ is computed directly with the same forward model (shown in Fig.~\ref{fig:Sketch_method}), by sampling 100,000 statistically independent atomic configurations from the quasi-harmonic approximation and averaging their FFT-based scattering intensities. 
The real-space extent of the supercell in our simulation is $L = 17.6$ \AA\, along each direction. The corresponding reciprocal-space discretization is therefore $\Delta q = 2\pi/L \approx 0.357$ \AA$^{-1}$. The TDS intensity is evaluated on a uniform $81\times81\times81$ grid in reciprocal space centered at $\mathbf{q}=\mathbf{0}$, spanning wavevectors from $-14.28$ \AA$^{-1}$ to $14.28$ \AA$^{-1}$  along each reciprocal-space direction. This grid provides a sufficiently dense and symmetric sampling of the diffuse scattering signal 
in the reciprocal space,
%
%around the Brillouin-zone center while 
capturing the relevant phonon-driven intensity modulations over multiple Brillouin zones.
%
%\hl{(TDS values saved over a cubic region in the reciprocal space, describe the range, describe the grid size.)}

%\hl{What we actually did is to use our own forward process (from EAM's IFC) to compute C-TDS rom 1000 configurations -- UPDATED in the text. (1) How about computing C-TDS using 10,000 or 100,000 configurations?  (2) How about computing C-TDS from MD snapshots?}

To learn IFC values from the target TDS signal, we use the workflow depicted in Fig.~\ref{fig:Sketch_method}.
As an initial guess, the values of the IFCs are set to those computed from a different interatomic potential of Ni, namely the modified embedded-atom method (MEAM) potential~\cite{baskes1997determination}.
The IFC values are optimized using the Adam optimizer for $6000$ epochs, and the learning rate is reduced by a factor of 2 from $0.4$ to $0.1$ every 2000 epochs.
During each epoch, we generated 15 atomic configurations from which we calculated an average scattering image for the fitting process to have a robust estimate of the TDS whilst limiting computational costs. Computations were performed on an NVIDIA A100 GPU (Ampere architecture) using CUDA 12.5. A single epoch executes in approximately 40~s on a single GPU, and convergence was typically achieved after 4000 epochs ($<$ 50 GPU hours), at which point the loss function has stabilized.

\subsection{Learning IFCs from Full TDS Data in 3D \textbf{q}-Space}
\begin{figure}[!htb]
    \centering
    \includegraphics[width=\linewidth]{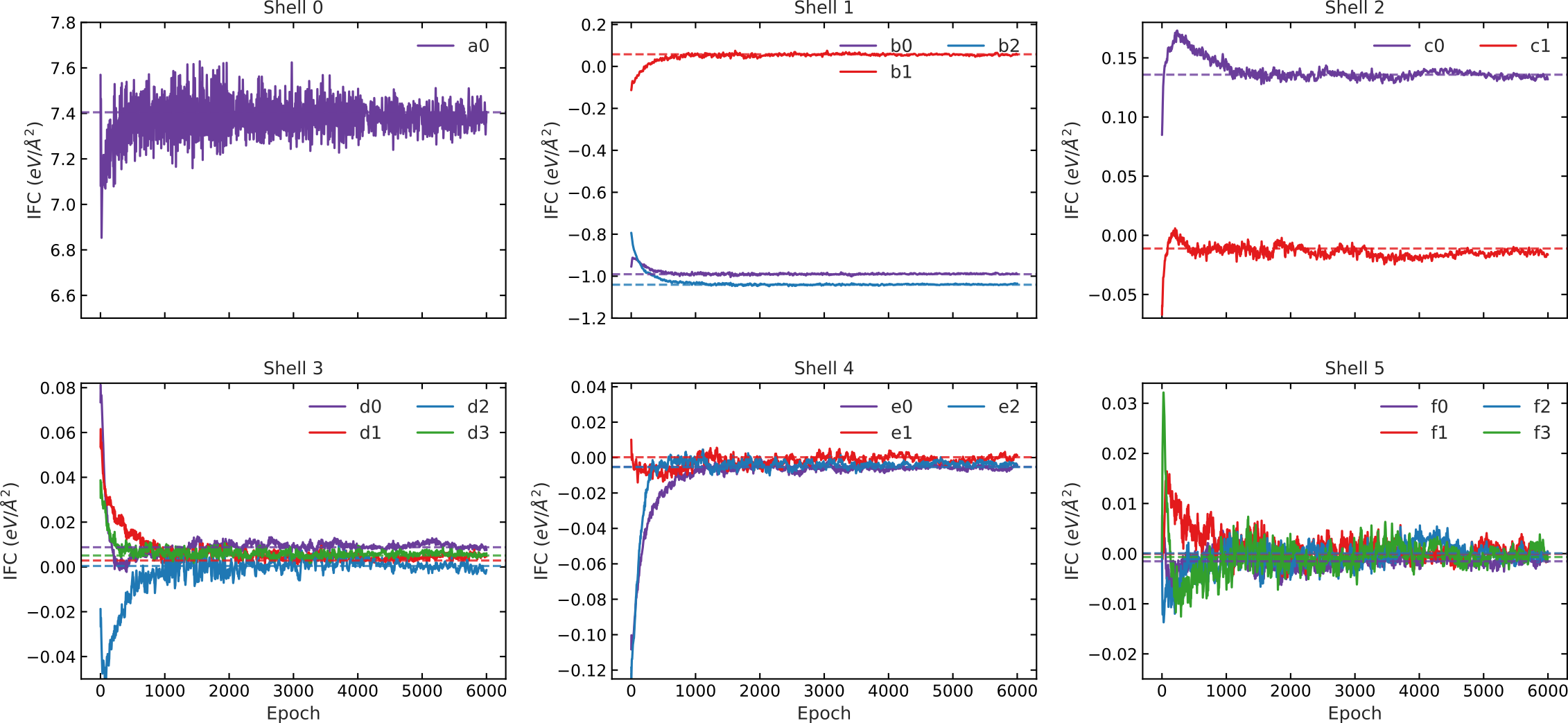}
    \caption{The evolution of the fitted IFCs corresponding to different neighboring shells during training.  The initial values of IFCs correspond to the MEAM potential, while the target values correspond to the U3 EAM potential.
    %for the individual shells show that after 6000 epochs, the force constants converge and closely resemble the target U3-EAM IFCs after fitting the initialized MEAM IFCs.
    %
    %
    %(3D data to fit, ref X-TDS from 1000 configs, 0-6000 epochs)
    }
    \label{fig:IFC_shells_converge}
\end{figure}
%

%Convergence of loss and Phonon dispersion relation
%
Fig.~\ref{fig:MEAM}(a) compares the TDS signal (on a 2D slice of $q_z = 0$) for the target EAM potential and the MEAM potentials used as initial guess; qualitatively the images look similar. However, the differences in the IFC values for the EAM and MEAM potentials lead to discernible patterns when the differences between their TDS predictions are plotted in Fig.~\ref{fig:MEAM}(b) (left half), as well as 
clear differences in the phonon dispersion relation shown in Fig.~\ref{fig:MEAM}(d).
Fig.~\ref{fig:MEAM}(c) shows the evolution of the loss function over the training period, demonstrating steady convergence towards the end of the training phase.
Already after 1000 optimization epochs, the loss in the simulated TDS intensity image converges to below $8\times10^{-4}$, indicating that the relative spectral shape of the scattering image is well described.
Fig.~\ref{fig:MEAM}(b) shows the spatial loss distribution of the TDS images across reciprocal space before and after optimization, confirming that the fitted IFCs substantially reduce the intensity mismatch over the full $\mathbf{q}$-space region.
The reconstructed phonon dispersion relation along the high symmetry path, shown in Fig. ~\ref{fig:MEAM}d, closely resembles the reference EAM dispersion. %The relative error is below $2\times 10^{-2}$ for all points along the path, and the absolute error is below $6\times 10^{-2}$.
The dispersion relations agree at points near the Brillouin center, only high frequency phonon modes at the Brillouin zone edge (e.g. at X, X*) exhibit any noticeable (but slight) deviation. This demonstrates that the approach enables accurate and precise reconstruction of the phonon dispersion across the Brillouin zone, consistent with reported results for comparable, non differentiable methods~\cite{holt_determination_1999, holt_phonon_2002, holt_x-ray_2001}.

%Convergence of IFCS
Fig.~\ref{fig:IFC_shells_converge} shows the evolution of the individual IFC values during optimization. The dominant first-nearest-neighbor shell ($b_k$) converges rapidly and stabilizes after approximately 1000 epochs, while the second- and third-nearest-neighbor shells converge more gradually. All IFCs converge to within the statistical uncertainty of the target values, with residual fluctuations reflecting the finite learning rate and sampling. %The fourth-nearest-neighbor shell is recovered with moderate accuracy, whereas the fifth-nearest-neighbor shell remains only weakly constrained by the TDS data and exhibits noticeably larger deviations from the reference values.  
Quantitative convergence errors for all shells are summarized in Table~\ref{tab:ifc_convergence_meam}. The absolute errors for force constants $b_0$ through $f_3$ are all on the order of $10^{-3}$. The $a_0$ parameter describing self-interaction, derived from the force constants according to Eq.~(\ref{eq:a0_force_constants}), exhibits an absolute error on the order of $10^{-2}$, though its relative error remains less than $10^{-3}$.
All parameters corresponding to the interaction of close neighbor shells 0 to 4 ($a_0$ to $e_2$) demonstrate progressive error reduction through the optimization progress. In contrast, fifth-shell constants ($f_1$-$f_5$) maintain relatively constant error magnitudes, indicating limited convergence improvement for these distant interactions.
This behavior reflects the decaying sensitivity of the diffuse scattering signal to increasingly distant interatomic interactions. While the dominant short-range force constants are accurately recovered, interactions beyond the fourth neighbor shell contribute only weakly to the scattering intensity and therefore cannot be reliably determined from TDS alone. 
%
%\hl{(Perhaps it is because we computed the reference C-TDS using only 1000 configurations?)}

The successful recovery of the dominant force constant corresponding to the close neighbor interactions confirms the effectiveness of the approach to recover the correct interaction potential even when starting from a different interatomic potential yielding very similar TDS images, emphasizing the power of quantitative TDS analysis over purely visual comparison.
We hereby show that, in addition to reproducing the phonon dispersion relation, the vibrational information contained in the scattering images allows for the precise extraction of the dominant IFCs, including interactions up to the fourth nearest neighbors.

\begin{table}[ht]
    \centering
    \caption{Comparison between IFC values in eV/\AA$^2$ for training using TDS data on the entire 3D grid of $\mathbf{q}$-space.  The ``Starting'' column corresponds to the IFCs at the beginning of the training (taken from the MEAM potential).  The ``Target'' column corresponds to the ground truth (here taken as the U3 EAM potential). The ``Converged'' column corresponds to the IFCs at the end of the training; its difference from the ``Target'' column is listed in the ``Error'' column.
    %Convergence of interatomic force constants after 10 000 configs (in eV/\AA$^2$).
    %
    %(3D data to fit, ref X-TDS from 10000 configs, after 6000 epochs) 
    }
    \label{tab:ifc_convergence_meam}
    \renewcommand{\arraystretch}{1.15} 
    \begin{tabular}{
        c
        S[table-format=-1.3e+1]
        S[table-format=-1.3e+1]
        S[table-format=-1.3e+1]
        S[table-format=-1.3e+1]
    }
        \toprule
        \textbf{} & \textbf{Starting} & \textbf{Target} & \textbf{Converged} & \textbf{Error} \\
        \midrule
        $a_0$ & 7.569 & 7.405e-0 & 7.406e-0 & -1.5e-3\\
        \midrule
        $b_0$ & -9.546e-01 & -9.907e-01 & -9.903e-01 & -3.480e-04 \\
        $b_1$ & -1.131e-01 & 5.773e-02  & 5.557e-02  & 2.155e-03 \\
        $b_2$ & -7.934e-01  & -1.041e+00 & -1.035 & -5.471e-03 \\
        \midrule
        $c_0$ & 8.487e-02  & 1.358e-01  & 1.318e-01  & 3.978e-03 \\
        $c_1$ & -6.711e-02  & -1.119e-02  & -1.6346e-02  & 5.160e-03 \\
        \midrule
        $d_0$ & 7.344e-02  & 8.787e-03  & 7.479e-03  & 1.308e-03 \\
        $d_1$ & 5.354e-02  & 2.866e-03  & 4.682e-03  & -1.816e-03 \\
        $d_2$ & -1.877e-02  & 4.220e-04  & -1.256e-03  & 1.678e-03 \\
        $d_3$ & 3.866e-02  & 5.112e-03  & 6.068e-03  & -9.563e-04 \\
        \midrule
        $e_0$ & -1.083e-01  & -5.230e-03  &  -4.565e-03  & -6.646e-04 \\
        $e_1$ & 1.009e-02  & 1.864e-04  & 1.214e-03  & -1.027e-03 \\
        $e_2$ & -1.184e-01  & -5.371e-03  & -4.391e-03 & -9.795e-04 \\
        \midrule
        $f_0$ & -1.975e-08  & -1.530e-03 & -1.660e-03  & 1.293e-04 \\
        $f_1$ & -1.975e-08  & -5.232e-05  & 3.958e-04  & -4.481e-04 \\
        $f_2$ &  -1.975e-08  & 6.624e-05  & 3.586e-04  & -2.923e-04 \\
        $f_3$ & -1.446e-21  & -6.790e-04 & -8.322e-04 & 1.532e-04 \\
        \bottomrule
    \end{tabular}
\end{table}

\subsection{Learning IFCs from Partial TDS Data in 3D \textbf{q}-Space}

%\hl{(Merge the following two paragraphs into one)}

In practice, acquiring TDS data across the full 3D reciprocal space typically requires sample rotation. Although full 3D TDS acquisition is now routine for ideal specimens, many experimental scenarios render sample rotation impractical or impossible. For instance, {\it in situ} monitoring of rapid phase transformations or the study of polycrystalline materials often prohibits sample rotation during observation, limiting data collection to sparse reciprocal-space coverage. Here, we investigate the feasibility of extracting interatomic force constants (IFCs) from TDS data when full 3D coverage is unavailable.
We find that fitting a single 2D detector image generally results in incomplete parameter recovery due to a fundamental symmetry limitation inherent to two-dimensional datasets passing through the Brilloun zone center ($\Gamma$).
%Merely fitting a single 2D detector image yields generally incomplete parameter recovery.
%Crucially, this limitation is intrinsic and cannot be overcome by choosing a specific slice orientation. Since an experimental detector image necessarily contains the origin, a single detector image acquired without sample rotation cannot provide sufficient information for complete IFC reconstruction.  
%The limitation arises from a fundamental incompleteness inherent to any two-dimensional dataset containing the Brillouin-zone center. 
%
At the $\Gamma$-point, 
%At $\Gamma$, 
phonon eigenvectors decouple into pure Cartesian displacement directions; consequently, a single reciprocal-space slice only constrains force-constant components that project onto phonons polarized within that plane.
Force components perpendicular to the slice remain mathematically unconstrained, necessitating multi-slice data for full determination of IFCs.

\begin{figure}[!htbp]
    \centering
    %\vspace{2in}
    \includegraphics[width=0.95\linewidth]{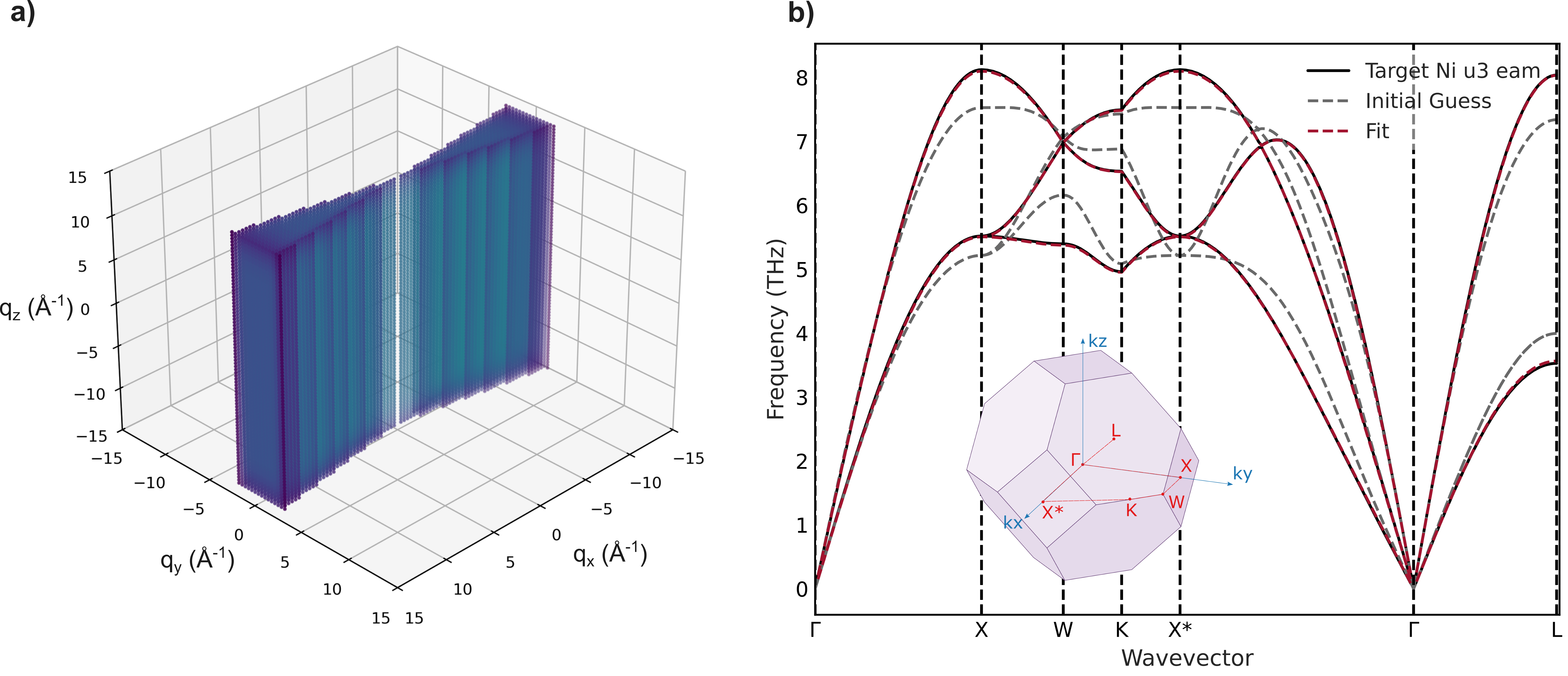}
    \caption{(a) A wedge in the reciprocal space.  Only the TDS data at $\mathbf{q}$ points inside the wedge are used to fit the IFCs. The wedge spans $\pm 10 \degree$ around the $q_y = 0$, a.k.a. the H(K=0)L, plane, allowing sampling of out-of-plane phonon modes.
    (b) The phonon dispersion relation computed from the Target, Initial, and Fitted IFCs using only TDS data from the wedge region. The curves corresponding to the Target and Fitted IFCs are nearly indistinguishable here. The inset shows the first Brillouin zone with the paths connecting high symmetry points shown in red.
    %The fitted IFCs can be shown to reproduce the phonon dispersion relation from the target U3 EAM accurately when only comparing the TDS on the 10 $\degree$ wedge. The inset shows the first Brillouin zone for, including the high symmetry path in red.
    }
    %\caption{(a) Comparison of the retrieved phonon dispersion relations for the full 3D TDS signal (green, dashed), the 2D HK0 slice (orange, dashed) and the 2D H,K,(L=H+K) slice (red, dashed) to the target (black). (b) Brillouin zone of fcc Ni, indicating the high-symmetry path along which the phonon dispersion relation is evaluated and the 2D reciprocal planes used for fitting. (c) Retrieved interatomic force constants compared to the target IFCs, highlighting the sensitivity of the fit to the specific reciprocal-space planes included in the data-set.\hl{(The two slices (HK0, HK(L=H+K)) do not show clearly.  Let's try to show them in a different way.  For each plane, find all the points where the plane intersects with the Broilluin zone polyhedron --- the result is a polygon.  Shade the polygon as a representation of the plane (slice).)}}
    \label{fig:phonon_wedge}
\end{figure}

\begin{figure}[!htb]
    \centering
    \includegraphics[width=\linewidth]{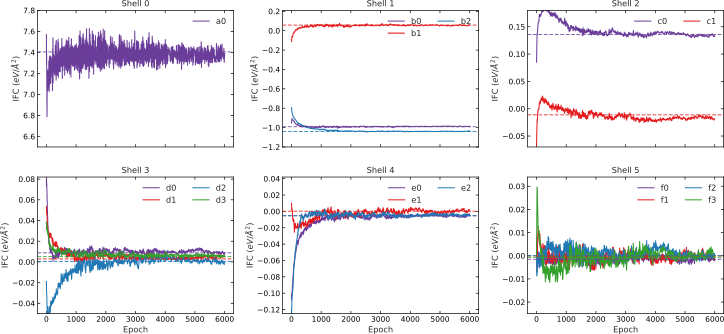}
    \caption{The IFCs for the individual shells show that after 6000 epochs, the force constants converge and closely resemble the target U3 EAM IFCs after fitting the initialized MEAM IFCs with a wedge of $\pm 10 \degree$ around the high symmetry HK0 slice ($q_z = 0$).
    %\hl{(Update this figure with the new result.  Caution: let's confirm this is indeed using the wedge but not the 3D fit.)}
    }
    \label{fig:IFC_shells_converge_wedge}
\end{figure}

A $\pm 10 \degree$ wedge provides a subset of $\mathbf{Q}$ which is a practical compromise between data completeness, fit-accuracy, and experimental feasibility. Experimentally, a double-wedge of $\pm 10\degree$ around a high symmetry plane, e.g. H(K=0)L can be accessed by rocking the angle between the sample and incoming X-ray beam. An idealized wedge is shown in Fig.~\ref{fig:phonon_wedge}(a). It should be noted that in the optimization procedure we apply all symmetry operations of the lattice (group ${\rm O}_{\rm h}$) to the computational TDS intensity, such that a $10 \degree$ wedge is equivalent to a larger fraction of the reciprocal space in a highly symmetric FCC lattice.
To assess if IFC recovery is possible on such an incomplete subset of $\mathbf{Q}$, phonon dispersion reproduction, and agreement of TDS intensity, are analyzed for IFCs obtained by fitting only to the $\mathbf{q}$ contained in the $10 \degree$ wedge.
Fig.~\ref{fig:phonon_wedge}(b) shows the reconstructed phonon dispersion relation along the high symmetry path. The reconstructed phonon dispersion closely resembles the reference EAM dispersion, with an accuracy comparable to the full three-dimensional fit. 
%Slight deviations from the EAM dispersion can be distinguished at the L point. While the path ($\Gamma$ - X - W - K) lies entirely within the HK(L=0) plane and is therefore well sampled, the $\Gamma$ - L direction points out of the plane and is hence only probed within the narrow  $10 \degree$ wedge. Since L is point furthest from $\Gamma$ in this out-of-plane direction, it is the least constrained by the reduced dataset, which can explain the slight deviations observed.

%

Fig.~\ref{fig:IFC_shells_converge_wedge} shows the evolution of the individual IFC values during optimization with 6000 epochs. The dominant first-nearest-neighbor shell converges rapidly and stabilizes after approximately 1000 epochs, while the second- and third-nearest-neighbor shells converge more gradually. The IFCs in the first-nearest neighbor shell converge to the target values within the statistical uncertainty. 
Quantitative convergence errors for all shells are summarized in Table~\ref{tab:ifc_convergence_meam_wedge}. The absolute errors for force constants $b_0$ through $f_3$ are all on the order of $10^{-3}$. The $a_0$ parameter describing self-interaction, derived from the force constants according to Eq.~(\ref{eq:a0_force_constants}), exhibits an absolute error on the order of $10^{-2}$, though its relative error remains at $2 \times 10^{-3}$.
All parameters corresponding to the interaction of close neighbor shells 0 to 4 ($a_0$ to $e_2$) demonstrate progressive error reduction through the optimization progress. In contrast, fifth-shell constants ($f_1$-$f_5$) maintain relatively constant error magnitudes, indicating limited convergence improvement for these distant interactions.
Similar to the full 3D dataset, the error on the individual IFCs for the $\pm 10 \degree$ wedge after 6000 epochs remain on the order of $10^{-3} - 10^{-4}$. Despite the comparably low values in the remaining error of the individual IFCs, the error in the self-interaction term  $a_0$ increases from $1.6 \times 10^{-3} $ to $1.5 \times 10^{-2} $. Since $a_0$ is derived from all IFCs, it serves as an overall measure of the fit quality.  Its significant increase, signifies an overall compromised fit quality. Despite this, the agreement of the individual IFCs and the phonon dispersion relation confirms that the wedge geometry captures the essential information needed for an accurate reconstruction. 
%Nonetheless, individual values for the second shell and above show significant deviations from the target values even after 6000 epochs,  particularly $c_1$, $d_2$ and $e_0$ and $e_1$. These force constants point in directions perpendicular to the H(K=0)L plane and are therefore not as fully constrained as in the 3D dataset. %The fourth-nearest-neighbor shell is recovered with moderate accuracy, whereas the fifth-nearest-neighbor shell remains only weakly constrained by the TDS data and exhibits noticeably larger deviations from the reference values.  

Our results therefore demonstrate that a sample rotation of $\pm 10 \degree$ provides a feasible path toward IFC refinement in cases where full sample rotation is not achievable, such as in {\it in situ} experiments.

\begin{table}[ht]
    \centering
    \caption{Comparison between IFC values in eV/\AA$^2$ for training using TDS data on a wedge region of the $\mathbf{q}$-space shown in Fig.~\ref{fig:phonon_wedge}(a).  The meanings of the columns are the same as in Table~\ref{tab:ifc_convergence_meam}.}
    %Convergence of interatomic force constants fitted to a wedge region in $\mathbf{q}$-space, using 100 000 configs for ref X-TDS (in eV/\AA$^2$).
    %
    %(3D data to fit, ref X-TDS from 10000 configs, after 6000 epochs)

    \label{tab:ifc_convergence_meam_wedge}
       \renewcommand{\arraystretch}{1.15} 
    \begin{tabular}{
        c
        S[table-format=-1.3e+1]
        S[table-format=-1.3e+1]
        S[table-format=-1.3e+1]
        S[table-format=-1.3e+1]
    }
        \toprule
        \textbf{} & \textbf{Starting} & \textbf{Target} & \textbf{Converged} & \textbf{Error} \\
        \midrule
        $a_0$ & 7.569 & 7.405 &  7.389& 1.602e-2  \\
        \midrule
        $b_0$ & -9.546e-01 & -9.907e-01 & -9.894e-01 & -1.311e-03 \\
        $b_1$ & -1.131e-01 & 5.773e-02  & 5.709e-02  & 6.413e-04 \\
        $b_2$ & -7.934e-01  & -1.041e+00 & -1.034e+00 & -6.130e-03 \\
        \midrule
        $c_0$ & 8.487e-02  & 1.358e-01  & 1.315e-01  & 4.199e-03 \\
        $c_1$ & -6.711e-02  & -1.119e-02  & -1.845e-02  & 7.262e-03 \\
        \midrule
        $d_0$ & 7.344e-02  & 8.787e-03  & 7.340e-03  & 1.388e-03 \\
        $d_1$ & 5.354e-02  & 2.866e-03  & 5.232e-03  & -2.367e-03 \\
        $d_2$ & -1.877e-02  & 4.220e-04  &-4.112e-04  & 8.223e-04 \\
        $d_3$ & 3.866e-02  & 5.112e-03  & 6.347e-03  & -1.235e-03 \\
        \midrule
        $e_0$ & -1.083e-01  & -5.230e-03  &  -4.252e-03  & -9.771e-04 \\
        $e_1$ & 1.009e-02  & 1.864e-04  & 1.687e-03  & -1.500e-03 \\
        $e_2$ & -1.184e-01  & -5.371e-03  & -4.919e-03 & -4.512e-04 \\
        \midrule
        $f_0$ & -1.975e-08  & -1.530e-03 & -2.144e-03  & 6.138e-04 \\
        $f_1$ & -1.975e-08  & -5.232e-05  & 3.088e-04  & -3.611e-04 \\
        $f_2$ &  -1.975e-08  & 6.624e-05  & 9.345e-04  & -8.682e-04 \\
        $f_3$ & -1.446e-21  & -6.790e-04 & -6.098e-05 & -6.180e-04 \\
        \bottomrule
    \end{tabular}
\end{table}

\section{Discussion} \label{sec:Discussion}

Our results demonstrate that thermal diffuse scattering contains rich information on lattice dynamics that can be systematically exploited to recover both phonon spectra and interatomic force constants. The differentiable framework developed herein enables the direct, quantitative extraction of this information, inaccessible by qualitative TDS analysis alone.
A direct visual comparison of the TDS images obtained from the target U3 EAM potential and the initial guess obtained from the MEAM potential (Fig.~\ref{fig:MEAM}a) illustrates a fundamental limitation of qualitative TDS analysis. Despite markedly different phonon dispersions (Fig.~\ref{fig:MEAM}d), the HK0 TDS slices of the two potentials are visually indistinguishable. The overall intensity distributions are similar, and the differences are subtle, amounting to a mean absolute deviation of only $2 \times 10^{-3}$. These deviations manifest as spatially distributed intensity variations rather than qualitative differences in the scattering pattern. Although the error in any individual feature is small, deviations accumulate coherently across reciprocal space, yielding a quantitative distinction that is statistically significant. This demonstrates that apparent agreement in TDS images does not necessarily imply correct phonon physics, and that qualitative inspection alone can be misleading. A quantitative framework is therefore essential for reliably validating interatomic potentials against diffuse scattering data.

% (we said it earlier)
%Benchmark tests on synthetic data of FCC nickel demonstrate that the approach accurately reconstructs phonon dispersion relations and BvK force constants from simulated TDS images, even when starting from perturbed or physically incorrect potentials.
%Furthermore, we find that while full 3D reciprocal-space coverage ensures the most accurate recovery, carefully chosen 2D slices, such as symmetry-appropriate planes, can yield nearly complete reconstructions.
%Importantly, the approach enables empirical discrimination between physically distinct interatomic potentials that produce similar lattice energies, and can iteratively refine incorrect potentials to recover the correct phonon-relevant force constants.

%Limitations and Outlook: 
The present framework operates within the harmonic approximation, which assumes that atomic displacements are small and that interatomic forces scale linearly with displacement. This approximation is accurate at low to moderate temperatures where thermal energies remain small. For the FCC nickel system studied here at 300 K, anharmonic contributions to TDS are negligible. At elevated temperatures, however, higher-order phonon–phonon interactions become significant, producing temperature-dependent phonon linewidths and frequency shifts that are especially prominent near phase transitions and cannot be captured by a static harmonic force constant matrix~\cite{barabash_diffuse_2009, xu_determination_2005}.
An important advantage of our approach is that, unlike eigenvalue-based methods, it is formulated directly in terms of atomic displacements. This provides additional entry points for incorporating temperature-dependent corrections and higher-order interactions through modified sampling schemes. In particular, normalizing flows can be used to sample atomic configurations directly from the interatomic potential~\cite{ahmad_free_2022}.  The resulting configurations are inherently anharmonic while the sampling procedure remains differentiable. Replacing the harmonic sampling scheme would, therefore, extend the framework to high-temperature regimes without sacrificing differentiability.

%Outlook: directly fitting IFCs to TDS data offers an atomistic validation for MLIPs
The ability to directly fit interatomic force constants to experimental TDS data is particularly relevant for the rapidly growing field of machine learning interatomic potentials (MLIPs) \cite{cui2023machine, rohskopf_empirical_2017, zguns_benchmarking_2025, han_benchmarking_2025}, whose accuracy depends critically on the representativeness of their training data. Conventional benchmarks like elastic moduli, phonon frequencies at high-symmetry points, or macroscopic thermal conductivity probe only bulk material properties and may fail to expose errors in the underlying potential energy landscape. Direct validation against experimentally derived IFCs, accessible through the present framework, would additionally provide an atomistic-level assessment of the fidelity across the full Brillouin zone. This positions TDS-based IFC extraction as a powerful complement to existing MLIP validation strategies, and suggests that diffuse scattering datasets could become a valuable addition to the training and benchmarking pipelines used in the development of next-generation interatomic potentials.

\section{Conclusion} \label{sec:conclusions}
%Our results show that the differentiable TDS framework efficiently recovers IFCs and phonon dispersion from X-TDS, even from perturbed potentials, from carefully chosen 2D TDS signal, and enables empirical validation and refinement of interatomic models.

%\hl{Perhaps keep the Conclusion section short, and move more detailed points to the Discussion section.}

We have demonstrated that interatomic force constants and phonon dispersion relations can be recovered directly from thermal diffuse scattering data through gradient-based optimization. The key enabler is a fully differentiable pipeline, from a symmetry-constrained Born–von Kármán parameterization of the Hessian, through Cholesky-based configuration sampling, to simulated TDS intensities. Applied to FCC nickel, the framework accurately recovers force constants up to the fourth nearest-neighbor shell from both full 3D reciprocal-space data and an experimentally realistic $\pm 10 \degree$ wedge domain, demonstrating robustness under partial data conditions, representative of experimentally constrained acquisition.
Basing the framework directly on atomic configuration space is not merely a computational convenience but makes the method extensible to anharmonic regimes and directly applicable to the validation of machine-learned interatomic potentials. Diffuse scattering data, long predominantly used qualitatively, can thus serve as a quantitative target for atomistic refinement.

\comment{We have developed a differentiable framework to retrieve interatomic force constants (IFCs) from thermal diffuse scattering (TDS) data, based on a Born–von Kármán (BvK) parameterization of the Hessian matrix.
By incorporating crystal symmetry constraints and a Cholesky decomposition of the Hessian, the method efficiently samples physically valid atomic configurations and computes TDS in a fully differentiable framework and thus enables gradient-based optimization using automatic differentiation.
The results show that TDS data contain sufficient information to recover not only the phonon spectrum but also the underlying interatomic force constants that govern lattice dynamics.
This demonstrates that gradient-based TDS fitting provides a powerful route for validating and tuning interatomic models directly against experimental scattering data.
In summary, this differentiable TDS fitting framework establishes a pathway toward experimentally grounded, phonon-aware refinement of interatomic potentials, bridging scattering measurements and atomistic modeling of lattice dynamics.}

\subsection*{Conflict of Interest}
The authors declare no competing interests.

\subsection*{Data Availability}
All data can be obtained on request from the corresponding author. The library for
the X-ray scattering calculations can be found at \href{https://gitlab.com/micronano\_public/c-xpcs}{C-XPCS}~\cite{Mohanty2022, mohanty2023evaluating}.

\subsection*{Acknowledgements}

K.S. thanks the Wenner-Gren Foundations (grant number WGF2023-0063).
This work is partly supported by the Air Force Office of Scientific Research
under award number FA9550-20-1-0397
(S.M.) and  FA9550-24-1-0237 (H.Z. and W.C.).

\renewcommand{\bibname}{References}
\addcontentsline{toc}{section}{References} %
\printbibliography
% \bibliographystyle{unsrtnat}
% \bibliography{references,refs}
\end{refsection}

% hide SI now to make it easier to finish this draft
%\newpage
%\input{supporting_information}

\end{document}